# A novel approach to formulae production and overconfidence measurement to reduce risk in spreadsheet modelling

S. Thorne, D. Ball, Z. Lawson
University of Wales Institute Cardiff

sthorne@uwic.ac.uk, dball@uwic.ac.uk, lawsonzf@cardiff.ac.uk

**ABSTRACT**

Research on formulae production in spreadsheets has established the practice as high risk yet unrecognised as such by industry. There are numerous software applications that are designed to audit formulae and find errors. However these are all post creation, designed to catch errors before the spreadsheet is deployed. As a general conclusion from EuSpRIG 2003 conference it was decided that the time has come to attempt novel solutions based on an understanding of human factors. Hence in this paper we examine one such possibility namely a novel example driven modelling approach. We discuss a control experiment that compares example driven modelling against traditional approaches over several progressively more difficult tests. The results are very interesting and certainly point to the value of further investigation of the example driven potential. Lastly we propose a method for statistically analysing the problem of overconfidence in spreadsheet modellers.

## 1. INTRODUCTION

In this paper we discuss one possible novel approach to formulae production and an overconfidence measurement to reduce risks in spreadsheet modelling. Our novel approach was the result of an inter-university collaborative project between UWIC and Cardiff Universities.

### 1.1 Spreadsheet errors

Since the widespread availability of the office PC there has been a rapid rise in End User Computing (EUC). EUC is a fairly broad term which covers database, presentation, word processing, spreadsheet and any other application that end users have access to. The activity of EUC was identified as presenting particular risks to organisations (Davis, 1987) (Alavi and Weiss, 1985) (Brown and Bostrom, 1989) (Munro et al, 1987) (Benson, 1983) (Brown and Bostrom, 1994) and (Alavi et al, 1987). Consequently, management strategies were formulated, designed to control EUC according to an organisations needs and culture (Brown and Bostrom, 1989) (Munro et al, 1987) and (Alavi et al, 1987).

Although spreadsheet modelling is incorporated into EUC management strategies, it requires special consideration due to the high flexibility, usability and availability that spreadsheets offer. For these reasons spreadsheets have become an indispensable tool to organisations for personal, strategic decision-making and even mission critical modelling. However, spreadsheet ease of use becomes problematic when organisations relying on spreadsheets in decision-making processes miss simple errors, which can then have serious repercussions. The EuSpRIG web site (www.eusprig.org) has some such examples: 'Cut and Paste error' cost Trans Atlanta Corp. $24 million (June, 2003), Seattle self storage company shares fell 7.1% after it was revealed they had unintentionally overpaid two chief executives $700,000 each (October, 2003) and Florida education executives duplicated the cost of an Elementary school by $12 million, seriously effecting the projects budget (September, 2003).

### 1.2 Human factors

Research has suggested that human factors influence the quality of spreadsheet models. Panko (1998) suggested that error rates in spreadsheets are similar to other areas of complex human cognition. Kruck (1998) proved that spreadsheet model quality could be improved by equipping modellers with the cognitive skills that they require to produce complex spreadsheets.

All of this research establishes spreadsheet use coupled with human factors as a high-risk activity for strategic decision-making in organisations. In order to lower the risk, spreadsheets and modeller interactions need to be examined in great detail and new alternative methods of interaction developed. In this regard, Panko at the 2003 EuSpRIG conference strongly encouraged the need for new approaches based on human factors research.

## 1.3 Example driven modelling (EDM)

Considering the existing research into spreadsheet error production, it seems that most of the problems arise from human cognitive errors. Panko (1998) discusses a basic error rate even for simple tasks and an increased rate for more challenging tasks. This begs further research into the precise nature of the relationships involved. As discussed below our research resulted in a negative exponential function using Halstead's 'complexity' measure (Halstead, 1977)

One interesting way to possibly improve the quality of spreadsheet models, taking account of the need for new approaches based upon human factor research, may be to eliminate the need for humans to produce formulae. Producing formulas is not what we are naturally best at (Michie, 1979). Human neural processes are mainly example driven and pattern matching. In contrast, computers are naturally good at arithmetic and logic (the Arithmetic Logic Unit is at the heart of a computer), but computers are not naturally good at generating examples or pattern matching. So perhaps at present we have things the wrong way round when producing spreadsheets, or at least not optimal.

For example, children learn arithmetic by example (Jon has four apples, Mary takes two. How many does Jon have left?), rather than solving: 4 - 2 = 2. When shopping at the supermarket, we know from previous examples when we are in danger of overspending. We rarely generate some formulae, which includes the effect of tomorrows interest rate rise on our credit card account, etc., and then use that formula to decide whether or not to buy those Garibaldi biscuits. Or again, we learn to catch a ball by example (i.e. practice) rather than dreaming up and solving the trajectory formulae. In this vein, we were interested in investigating whether it may be easier to provide simple examples, which satisfy the problem rather than derive the requisite formulae.

### 1.4 Overconfidence

Panko (2003) extended his human factors research to measure overconfidence in modellers and the subsequent effect it has on the spreadsheet data integrity. Panko found that overconfidence in modellers ranged from 80% to 100%. These results show that overconfidence is a significant issue in spreadsheet modelling. Overconfident modellers may fail to apply any methodology or testing strategy and will not question the validity of their model. This practice is clearly counter-productive and contributes to the current poor integrity of spreadsheet models, hence our interest in establishing a satisfactory metric. Our results extend research on overconfidence measurement.

## 2. EXPLORING FORMULAE PRODUCTION

### 2.1 Introduction

To examine how pervasive spreadsheet errors are in formulae production, in the face of increasing formula complexity, experiments were designed to establish error rates in both traditional formula production and example driven modelling (EDM). Traditional modelling was used as a control for the EDM experiments. The results of these different paradigms are then compared and analysed accordingly.

### 2.2 Aim

The aim of the experiment was to establish experimentally within an academic environment, using undergraduate and postgraduate students:

1. The relationship between spreadsheet error rate and formula complexity using a) Traditional modelling, b) EDM
2. The (hypothesised) superiority of EDM over traditional modelling.
3. A more satisfactory statistical measure of overconfidence.

### 2.3 Experiment design

#### 2.3.1 Introduction

This experiment was designed in accord with Campbell and Stanley (1963), which is considered a seminal text in Quasi and Experimental research design. Our experiment randomly selected a large target group (57 students) from a universe (of 2000 university students). The necessary experimental details were established using a pilot experiment using a smaller sample (12 students).

#### 2.3.2 Sampling

Participant selection is critical to the credibility of the experiment. In order to minimise bias of inexperience, certain courses were targeted. Considering similar studies: (Panko and Sprague, 1998); (Panko and Halveson, 1996); (Galletta et al, 1997); (Galletta et al, 1993); (Teo and Tan, 1997); (Panko and Halverson, 1997) and (Irons, 2003) these all used either undergraduate students or Masters level students as participants. For comparable results several different courses were targeted, which also maximised the number of results. The courses targeted were: Final year Undergraduate Business Information Systems and MSc Information Systems. These groups have been selected for their academic and industrial experience, other courses were deemed to be unsuitable for testing. Participants were then selected randomly from these groups.

The participant's previous experience in spreadsheet development varies. All participants have, at some point, undertaken a module that focuses on EUC development packages (Including spreadsheets with a specific assignment). Students are exposed to spreadsheets all through their university life, they are used in business type modules but also in statistics modules. It is a fair assumption that the participants also use spreadsheets outside of university life in some capacity. All participants would have at least a basic working knowledge of spreadsheets and would have been exposed to creating spreadsheet formulae. It is likely that the participants would have tackled similar problems. To further ground the experience level, all participants were given a brief lecture and document detailing the construct of various spreadsheet formulas.

Equal importance should be given to sample size, considering the studies mentioned above, the average number of participants was 52. In our case the number of participants was 57.

#### 2.3.3 The Experiment

Students from the relevant groups were given a series of 5 tasks to complete using two different approaches. The tasks involved a 'traditional' approach of manually constructing spreadsheet formulae (serving as a control) and an EDM approach (where the participants were required to give example data for various attribute classifications)

The experiment was conducted in three stages. The first was to deliver a brief reminder lecture on how to use formulae in excel and some practical demonstrations. The second stage was a series of five tests involving formulae production. This was followed by a self-evaluation of perceived accuracy that was compared to the actual accuracy to measure over/under confidence. The third stage tested EDM by repeating the first five tests, but instead of creating formulae, the participants were required to give correct example data in accord with the model. This was again followed by their self-evaluation of perceived accuracy that was compared to the actual accuracy to measure over/under confidence. The tests in both cases were progressively more challenging, each building on the difficulty of the last.

#### 2.3.4 Confidence questionnaire

As stated, once the questions were completed the participants were required to fill in a short self-evaluation of perceived accuracy questionnaire on 'task completion' and 'confidence measure'. This measured how confident the participants were and was later compared with the actual result.

#### 2.3.5 Measuring the dependent variable and error production metrics

For a spreadsheet formulae the dependant variable's success or otherwise is determined when the formula that is produced has the correct syntax and has the correct cell referencing. To put it simply, where mistakes were not obvious, the formulae were inserted into a known populated spreadsheet to check their validity by inspection. EDM success was determined by inserting the answer values into known working formulae to see if the data was valid. The resulting formulae were examined and the 'percentage of models with errors' (Panko, 1998), 'percentage accuracy' and 'task complexity' deduced.

The questions were measured in terms of Halstead's complexity (Halstead, 1977). Halstead was used because it is a generally accepted metric and uniquely not only considers software complexity but also within formulae production. Using Halstead's complexity test (Halstead, 1977), the questions were given a value allowing relative comparison. Halstead's test was originally developed as a means of determining the relative complexity of software and algorithms within software and formed part of a larger set of tests to determine volume, difficulty, effort and complexity. Halstead's complexity formula, which is commonly referred to as *Halstead's complexity,* is shown below. In spreadsheet formulae, the operands and operators are determined across the whole formulae. Operators include: IF; AND; NOT; AVERAGE; SUM; >; = etc. Operands are the cell references and numbers used in the formulae. The resulting *complexity* ranges from 0, being the most complex, to 2 being the least complex.
where

n1 = the number of distinct operators
n2 = the number of distinct operands
N1 = the total number of operators
N2 = the total number of operands

$$Complexity = \frac{2*n1}{n2*N2}$$

The percentage accuracy for the sample group was then plotted against the Halstead complexity.

### 2.3.6 Overconfidence measures

As there are no existing satisfactory statistical measures of overconfidence, mainly attributed to its relative newness in the field, a method has been created to achieve this goal. The 'overconfidence ratio' is a coefficient value ranging from 1 (worst) to 5 (perfect). This value matches the participants 'combined overconfidence' value to the actual result F(x) by questions 1 to 5.
The combined overconfidence is created by using the questionnaire results for confidence (1 to 5) and perceived difficulty (1 to 5). Hence:

Combined overconfidence = (A*Confidence) + (B*Difficulty)
where A = B = 0.5

Define X as the number of errors, F(X) = mark/actual result

| | | |
|---|---|---|
| F(x) = | 5 | if x = 0 |
| | 4 | if x = 1 |
| | 3 | if x = 2 |
| | 2 | if x = 3 |
| | 1 | if x >= 4 |

(F(x) = 0 if the question was not attempted)

$$Confidence \quad \text{ratio} = \frac{\text{Ratio percieved error rate}}{\text{Actual error rate}}$$

In real terms this measures the expected outcome of the participants to the actual outcome and ranks it according to how accurate their prediction was, resulting in a value of 1, indicating a perfect match between expected and actual, to 5 indicating the worst match between expected and actual. It may also range below 1 to 1 fifth where greater than 1 implies over confidence and less than 1 implies under confidence

### 2.3.7 Participant tasks

The participants were given 5 descriptive problems based around creating formulae to produce grades for a hypothetical set of marks (no actual data was included) for a university. It was then up to the participants to produce a formula that would solve the problem. The questions got progressively more complicated, requiring the participants to account for different factors in the formulae. For example, the first question required a formula that would output “pass” or “fail”. Later questions required the formula to output “Fail”, “Compensate”, “Pass”, “Merit” and “Distinction”. The participants were also required to use indirect referencing and variable grade boundaries as the questions progressed.

For example one possible solution to “Ensure both exam and coursework are above 40 and can have the classification: Fail (<40); Pass (>=40, <55); Merit (>=55, <70) or Distinction (>=70) ” formulae is: -

=IF(MIN(C5:D5)<40,"Fail",IF(AVERAGE(C5:D5)>=70,"Dist",IF(AVERAGE(C5:D5)>=55, "Merit",IF(AVERAGE(C5:D5)>=40,"Pass"))))

The traditional formulae success or otherwise was determined on whether the formula that was produced had the correct syntax and the correct cell referencing. To put it simply, where mistakes were not obvious, the formulae were inserted into a known populated spreadsheet to check their validity by inspection. EDM success was determined by inserting the answer values into known working formulae to see if the data was valid. The confidence questionnaire then followed this.

As the second part of the experiment, the participants were given the same descriptive problems, as questions 1 to 5, but instead of producing formulae, they were required to give example data for each classification. The second confidence questionnaire then followed.

### 2.3.8 Experimental details

Described below are the conditions and details for the experiment conducted in an academic arena.

#### Conditions

There was no conferring allowed between by participants; each test was unique to a participant. There was no time limit imposed on the test, the test did have to be completed in the presence of an examiner. The participants were not told the nature of the experiment.

#### Details

The participants were given two documents for the test. The first document was designed to accompany the brief reminder lecture given to the participants. The second document was the question, answer and confidence papers. The participants were talked through the documentation they had been given and were told where to write the answers and what to do with the papers when finished. They were then allowed to start the test.

## 3. ANALYSIS

The results that this experiment provided are broken down into two sections. The first deals with the accuracy of manually modelling formulae against the example driven approach and the second deals with overconfidence in modellers.

### 3.1 Traditional approach

The results from the traditional modelling show a high error rate. The average percentage of incorrect answers for questions 1 to 5 was 80%. The percentage of models with errors, Panko (1998) is therefore 80%. This high error rate is typical of similar studies such as: 100% (Hicks, 1995); 91% (Coopers and Lybrand, 1997); 91% (KPMG, 1997); 86% (Butler, 2000); 84 and 95% (Janvrin and Morrison, 1996 and 2000) and 80% (Panko and Halverson, 1997).

Figure 1 shows the percentage accuracy drops with increasing Halstead complexity, as the questions progress. The traditional method also yielded an average of 4 mistakes per question. EDM, in contrast gave an average of 0.3 mistakes per question.

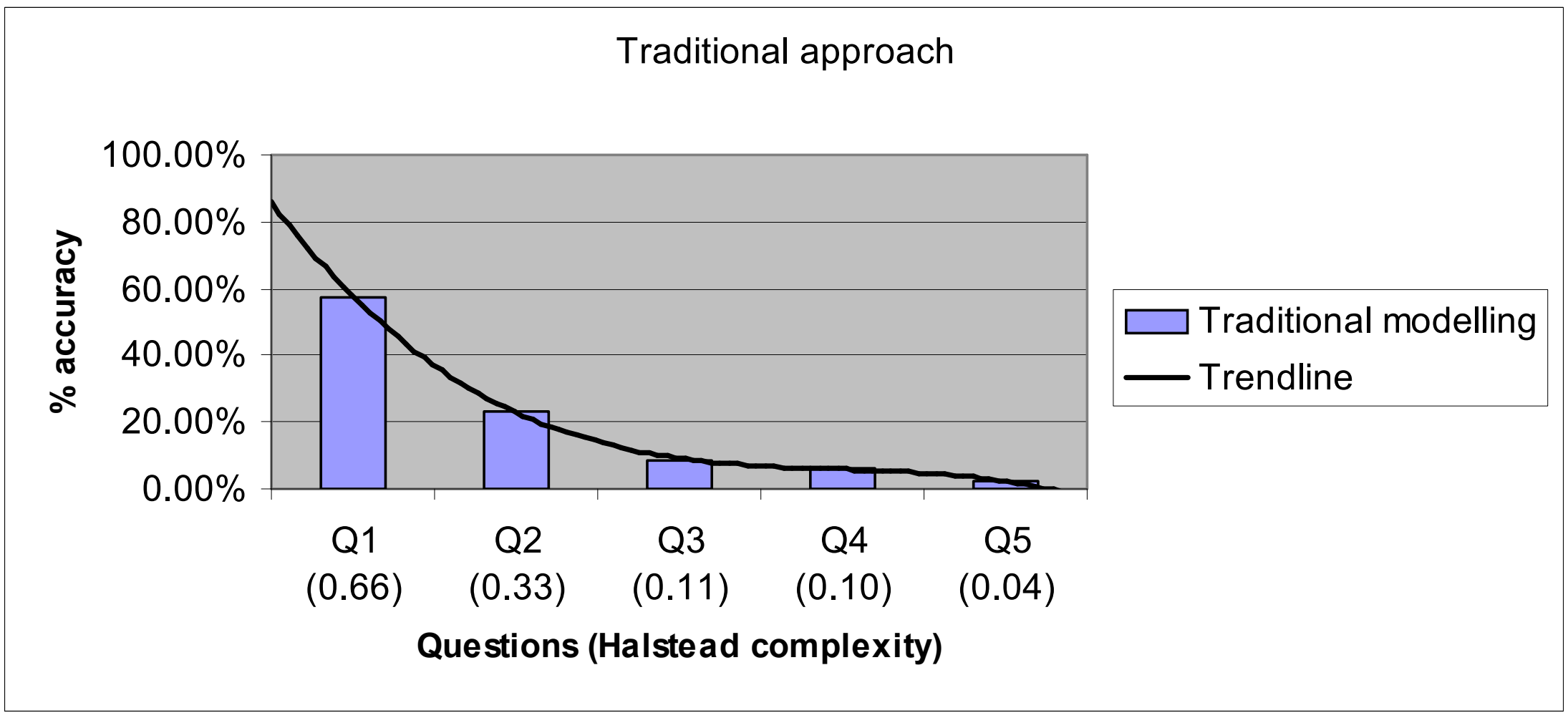

**Figure 1**

The curve in figure 1 appears to be exponentially decreasing; it does however demonstrate some strange behaviour at either extremity. We believe that the equation is trying to cater for the boundary conditions at either extremity, see figure 2.

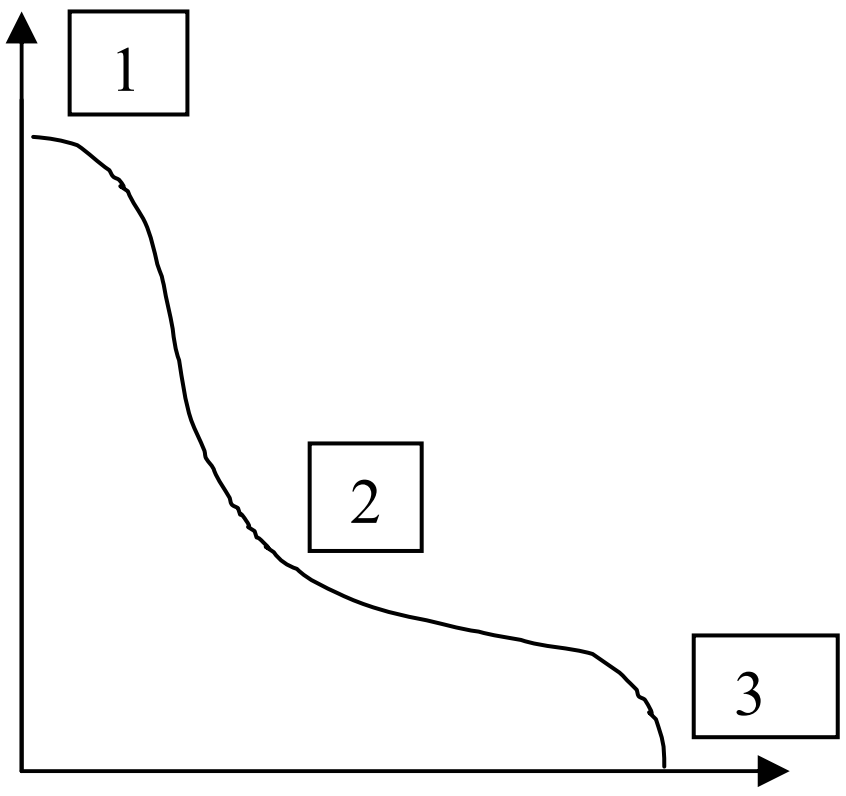

**Figure 2**

In the case of very low complexity, there are good reasons for believing that the percentage accuracy will never reach 100%. This upper boundary is due to the base error rate (Panko, 1998) and therefore a limitation occurs and the curve plateaus. See figure 3.

In the case of very high complexity, we believe that a similar limitation occurs and the curve becomes a vertical line. The reason for this kink is due to the well-known constraints of human working memory (Miller, 1956) which states that human memory (of the order of a few minutes) is restricted to handling ‘seven plus or minus two concepts simultaneously’. For example nested IF statements that include AND, averaged cells with indirect referencing is starting to get quite complex. When the number of concepts being handled simultaneously exceeds 9 (the Miller threshold), then unless there is some kind of spreadsheet engineering technique used, errors are almost certain. This point is crucially important when as far as mission critical spreadsheets are concerned. It would be very interesting to know what percentage of mission critical spreadsheets have formula beyond the Miller threshold, where also such organisations have no technique for coping beyond the Miller threshold, i.e. a spreadsheet modelling methodology or equivalent.

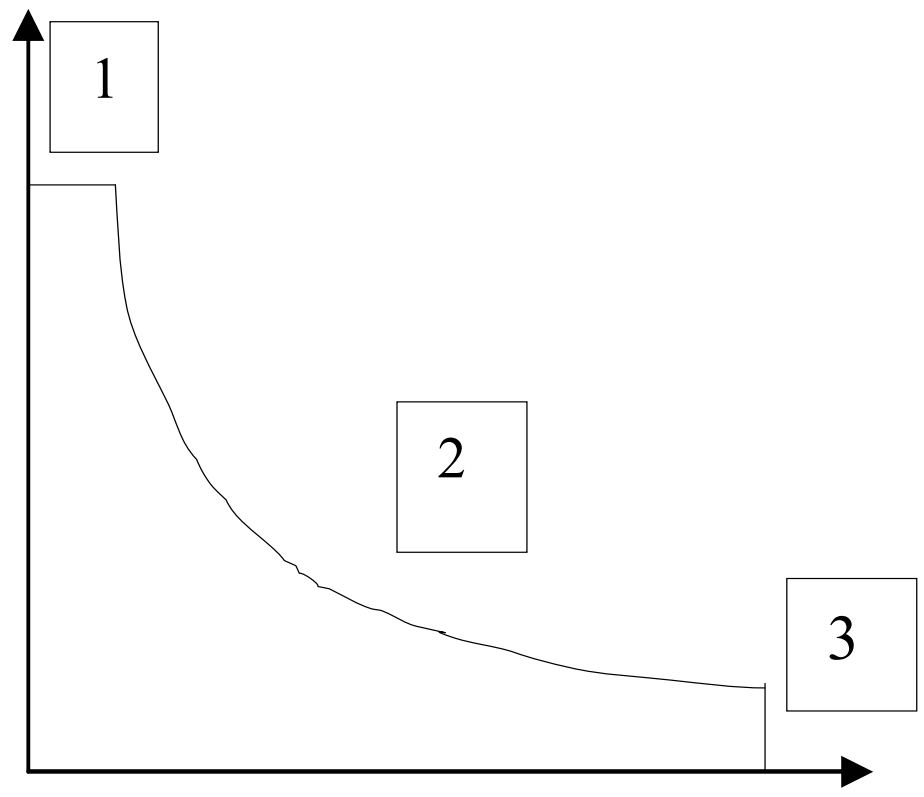

**Figure 3**

### 3.2 EDM

Figure 4 displays the results for EDM. As can be seen from the graph, the level of accuracy is very high and is easily able to cope with the most complex question. On average, 98% of all answers given were correct.

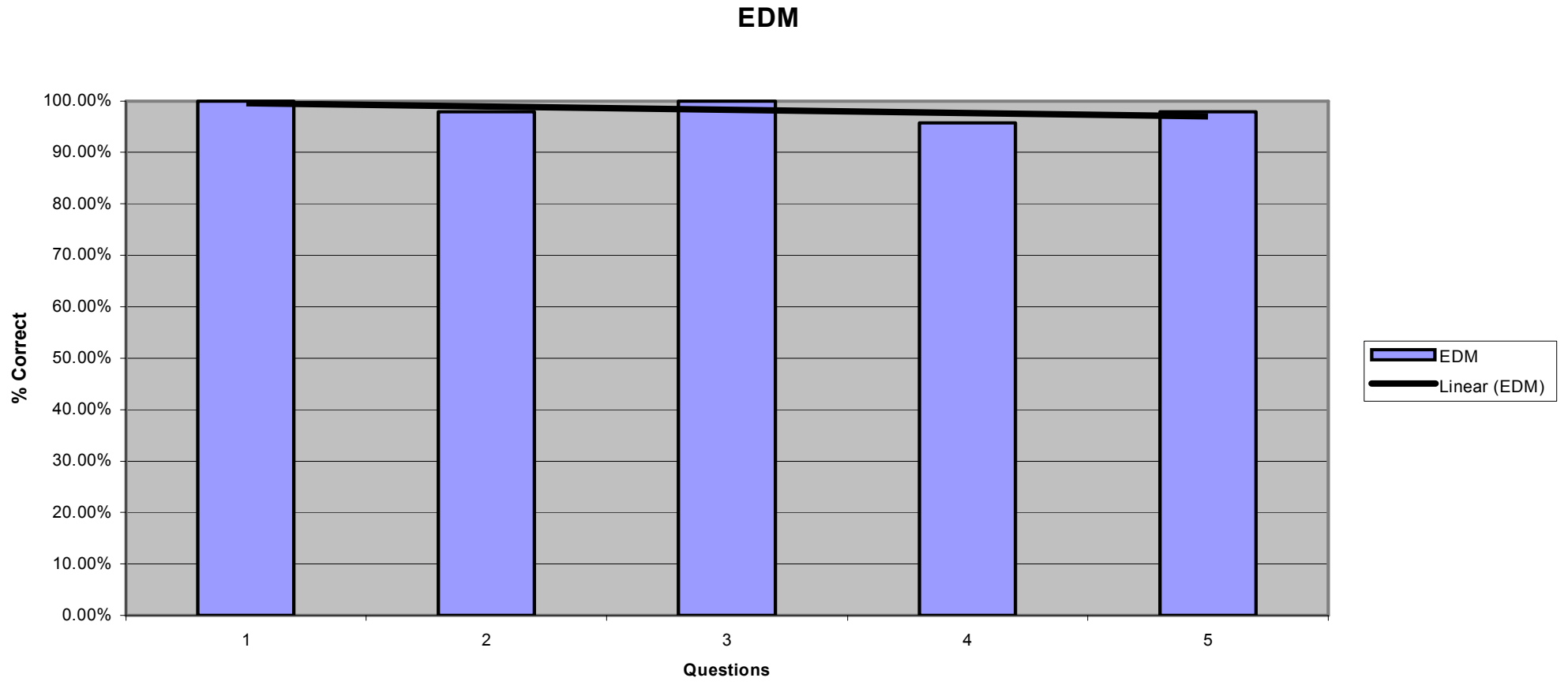

**1.1. Figure 4**

### 3.3 EDM and Traditional comparison

Figure 5 compares the difference between EDM and traditional methods. These results show a very significant improvement in accuracy when using an EDM approach. However, further research on EDM may well reveal new sources of error (Fraser and Smith, 1992)

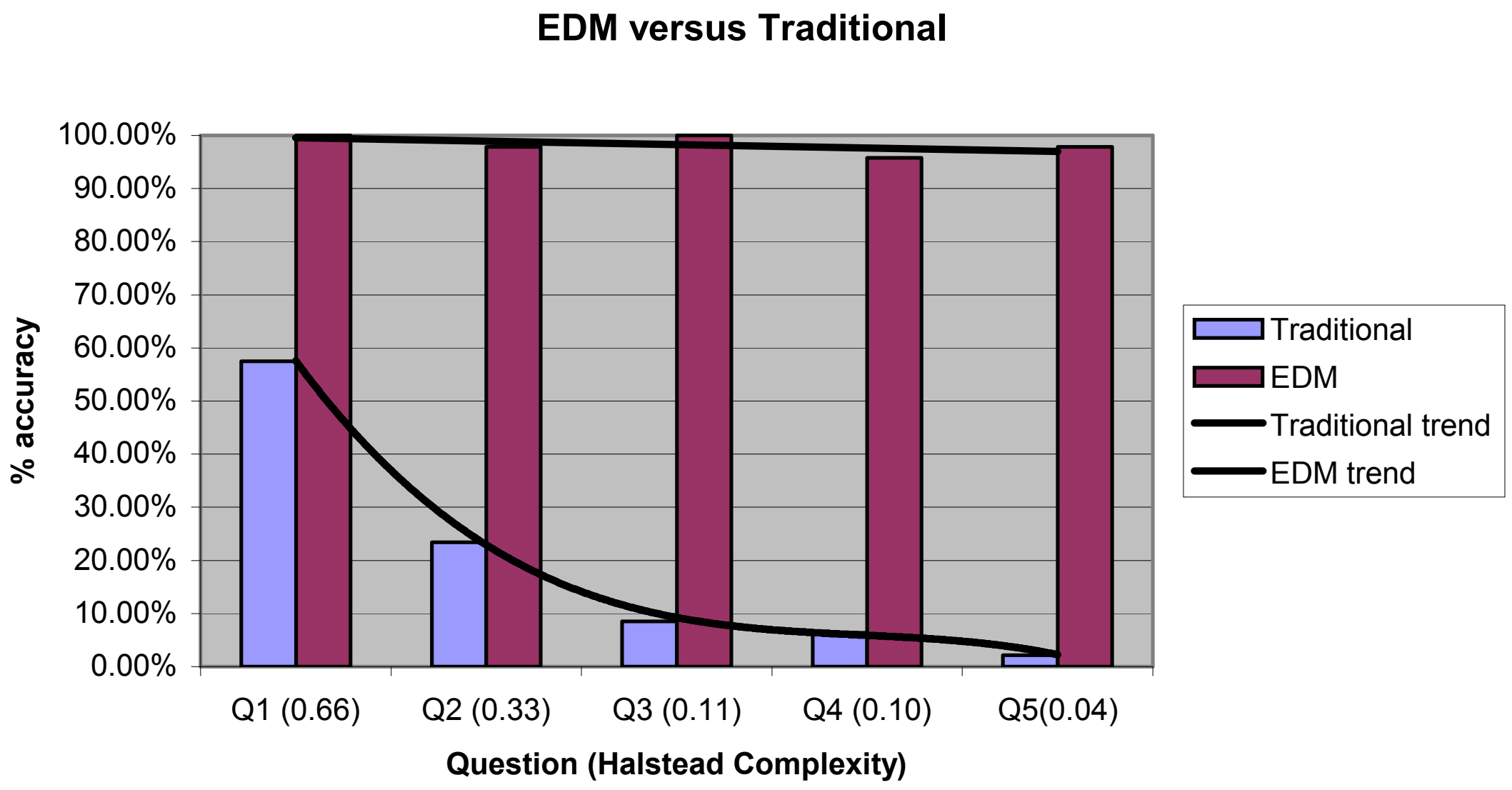

**Figure 5**

### 3.4 Overconfidence

Figure 6 shows the results for the confidence metric. Clearly the traditional approach leads to overconfidence. Our method for measuring confidence is based on how well the participants match their actual and perceived accuracy in assessing the difficulty of the task. A confidence ratio of 1 means a perfect match. It would appear that, within experimental error, this is true for EDM responses because the line starts just below and ends just above the confidence ratio of 1.

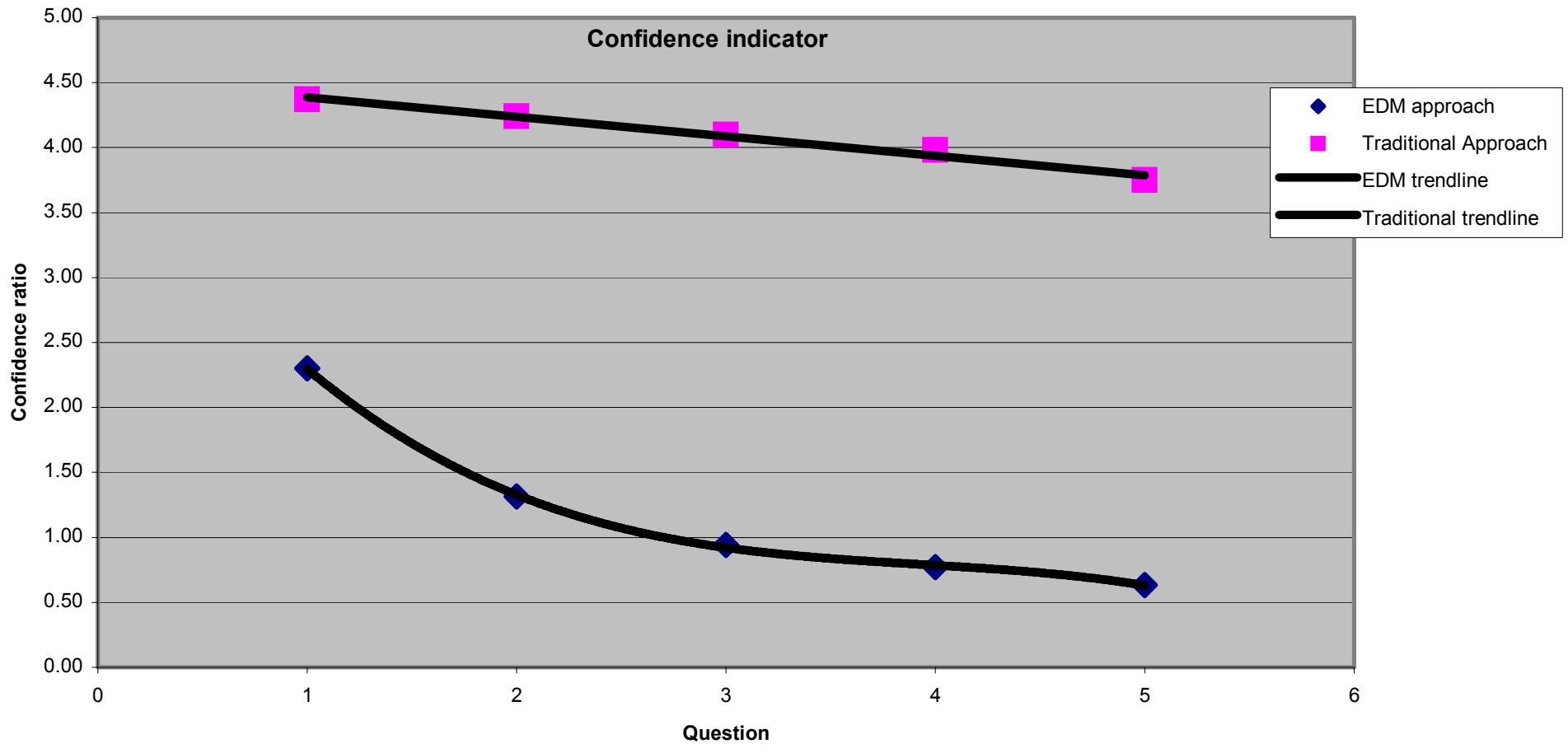

**Figure 6**

The questionnaire used in the experiment also allows relative comparison of perceived difficulty in the traditional and EDM approaches see figure 7. Note that, 1 corresponds to impossibly hard, 5 corresponds to trivially easy. The interesting point here is not the slow increase in perceived difficulty as the questions increase in complexity, since this was expected. Rather, the interesting point is the gap between the traditional and the EDM approaches. The gap corresponds to a significant usability advantage for EDM.

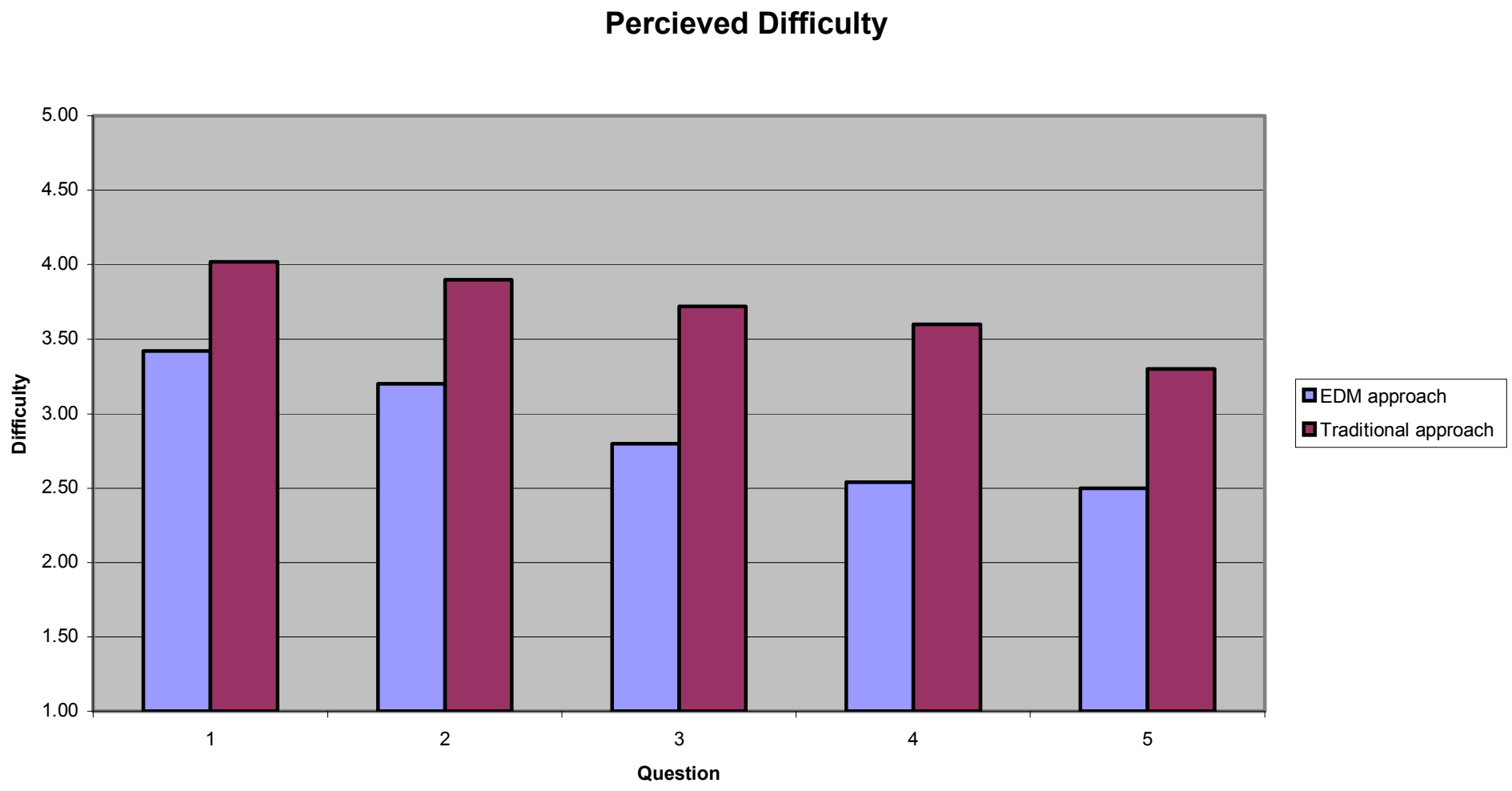

**Figure 7**

## 4 CONCLUSIONS

### 4.1 Validation of aims

The first aim was to establish the relationship between spreadsheet errors and formula complexity using traditional modelling and EDM. The relationship is a negative exponential but the model breaks down at both very high (Miller threshold) and very low (Base error rate) complexities.

The second aim was to consider whether EDM was superior to traditional modelling. As discussed above, the evidence certainly suggests this.

The third aim was to establish a more satisfactory measure of overconfidence. We feel our overconfidence measure has more transparency than existing methods. Further, relative comparisons between the traditional and EDM were more easily highlighted.

### 4.3 Limitations to the experiment

The experiment has been conducted in an academic environment; other environments also need to be considered. The EDM experiment suggests that there may be considerable merit in EDM being much more thoroughly investigated.

### 4.4 Further research

It would be very useful to verify the above findings within a real-life business-modelling environment. As discussed above it would be very interesting to assess the percentage of mission critical spreadsheets which contain formulae beyond the Miller threshold which do not make use of a rigorous spreadsheet engineering methodology or equivalent. Further research on EDM might include a machine to read in the examples and convert them to formulae. Many questions might be asked concerning the validity and application of such a machine, but this research points to possible advantages of such a novel advance.